\documentclass[runningheads]{llncs}
\usepackage[T1]{fontenc}
\usepackage{graphicx}
\usepackage{comment}
\usepackage{amsmath}
\usepackage{csquotes}
\usepackage{xcolor}

\usepackage{tikz}
\usetikzlibrary{calc}

\usepackage[colorinlistoftodos,obeyFinal]{todonotes}
\usepackage{booktabs}
\usepackage{subcaption}
\usepackage{float}
\usepackage[hidelinks]{hyperref}
\usepackage{xurl}    
\usepackage{array}
\usepackage{multirow}
\usepackage{pdflscape}
\usepackage{ragged2e}
\usepackage{placeins} 
\usepackage{fontawesome5} % icons
\newcommand{\orcidlink}[1]{%
  \href{https://orcid.org/#1}{{\color[HTML]{A6CE39}\faOrcid}}%
}

\usepackage{tabularx}
\usepackage{array}
\newcolumntype{Z}[1]{>{\raggedright\let\newline\\\arraybackslash\hspace{0pt}}p{#1}}
\newcolumntype{L}{>{\raggedright\let\newline\\\arraybackslash\hspace{0pt}}X}
\newcolumntype{R}{>{\raggedleft\arraybackslash}X}

\title{New Lenses on Learning Design: Course Types and Indicators from Large-Scale Moodle Data}

\titlerunning{New Lenses on Learning Design}

\author{Ekaterina Soroka\orcidlink{0009-0008-7783-1757} \and
Niels Seidel\orcidlink{0000-0003-1209-5038} \and
Marcus Specht\orcidlink{0000-0002-6086-8480} \and  \\
Kamila Misiejuk\orcidlink{0000-0003-0761-8703}}

\authorrunning{E. Soroka et al.}
\institute{CATALPA, FernUniversität in Hagen, 58097 Hagen, Germany \\
\email{\{ekaterina.soroka, niels.seidel, marcus.specht, kamila.misiejuk\}@fernuni-hagen.de}
}

\begin{document}

\AddToHook{shipout/foreground}{%
  \begin{tikzpicture}[remember picture,overlay]
    \node[opacity=0.2,scale=8,rotate=45] at (current page.center) {PREPRINT};
  \end{tikzpicture}%
}

\maketitle   

\begin{abstract}
Although learning design (LD) plays a central role in shaping online courses, large-scale empirical evidence on how courses are actually structured remains scarce, and existing analytics work has characterised courses through student engagement with specific learning activities rather than through indicators of the design itself. This study addresses both gaps through an analysis of learning management system (LMS) data from 606 courses at a distance university. Using LMS data, we first applied dimensionality reduction and clustering to a structural variable set, from which three distinct LD types emerged empirically: \textit{Compact}, \textit{Expansive}, and \textit{Interactive} and statistically characterised how their design elements differ across disciplines, activity load, and interfaces. Building on this typology, we derived three LD indicators, \textit{activity-mode balance}, \textit{structural clarity}, and \textit{analytics readiness}, each combining multiple LMS-derived variables into higher-order constructs that target a different aspect of LD. All three indicators differentiated significantly across the LD types and added pedagogically meaningful dimensions for interpretation. These contributions advance the operationalisation of LD in technology-enhanced learning research by demonstrating that higher-order LD indicators can be derived directly from routinely collected LMS data, and offer empirical insight into how LD manifests at scale within a large distance university.

\keywords{learning design \and learning analytics \and learning management system data \and indicators \and design analytics \and higher education}
\end{abstract}

\section{Introduction}

How instructors design their courses shapes the conditions for student learning. The pathways they create, the activities they sequence, and the ways they balance content, assessment, and collaboration all structure the environment in which student engagement unfolds. These decisions are increasingly enacted through learning management systems (LMS), where course structure, activity composition, and resource organisation are directly configurable by instructors. Particularly in fully online settings, these design decisions, collectively referred to as learning design (LD) \cite{koper2005learning}, mediate nearly every interaction between students and content, peers, and instructors. Without understanding how student learning is structured, it is not possible to explain why the environment affords or constrains particular behaviours, which is a central concern for technology-enhanced learning (TEL) research \cite{kirschner2015we}. 

This paper addresses that need through a large-scale analysis of LMS data from 606 courses at a large German distance university. We examine how recurring course structure patterns from LMS data can be identified as LD types, operationalised, and interpreted and how these insights can be used to develop meaningful LD indicators. In particular, we address the following research questions: \textbf{RQ1} What LD types characterise Moodle course designs at scale, and how do they differ in their characteristics? \textbf{RQ2} How can the LD indicators be derived from LMS course data, and how can derived LD indicators be used to characterize the LD types?

The contributions of this paper are fourfold: (1) a comprehensive large-scale analysis of course designs from Moodle data; (2) the identification of three distinct LD types from authentic institutional data using dimensionality reduction and clustering techniques; (3) a novel approach to derive LD indicators from both theoretical considerations and the analysis of LMS structural data; and (4) the introduction of three new LD indicators grounded in this approach.

\subsubsection{Background}
LD refers to the deliberate structuring of learning opportunities, encompassing decisions about activity types, sequencing, resources, and the social organisation of learning \cite{rienties2017review}. While LD encompasses a rich set of decisions including learning outcomes (LOs), intended task duration, or pedagogical strategy \cite{law2020multilevel}, LMS data provides a practical and scalable window into the enacted design, capturing the activity structures and resource configurations that instructors have made visible to students. Learning analytics (LA) offers the possibility of examining how these LD decisions play out in practice, by capturing and analysing traces of students' interactions with the learning environment \cite{schmitz2017opportunities}. \cite{hernandez2019analytics} introduced \textit{design analytics} as a distinct analytical perspective focused on the properties of a course as designed, rather than as experienced by students. Where learning analytics captures what students do, design analytics captures what instructors have put in place: goals, tasks, social configurations, temporal structures, and resource arrangements, as well as the design process, which can include task structure, workload balance, or implementation sustainability.

Empirical large-scale analyses of actual LDs remain scarce. The most substantial body of work comes from the Open University UK (OU UK), where the LDs of all university courses were coded and examined across seven faculties and multiple years \cite{nguyen2017mixing}. \cite{whitelock2020disciplinary} applied epistemic network analysis to 6,040 Moodle course offerings across four years and ten faculties to examine disciplinary differences in blended learning activity patterns. \cite{misiejuk2023changes} similarly used epistemic network analysis on 102 courses over three years at a Norwegian university to examine how Canvas use profiles shifted in response to the COVID-19 pandemic. 

The classification of learning activities into taxonomic categories has provided the dominant framework for characterising LD in LMS data. The most widely applied is the LD taxonomy developed at the OU UK, which distinguishes seven activity types: \textit{assimilative}, \textit{finding and handling information}, \textit{communication}, \textit{productive}, \textit{experiential}, \textit{interactive/adaptive}, and \textit{assessment} \cite{rienties2017review}. In practice, LMS activities are mapped onto these categories, for example forums and chats are classified as communicative, assignments and quizzes as assessment, and pages and resources as assimilative \cite{nguyen2017mixing}. This can be compared to a more granular taxonomy used by \cite{hernandez2019analytics},  further subdeviding these broad categories into six distinct learning activity types: \textit{acquisition, inquiry, practice, production, discussion,} and \textit{collaboration}. A similar categorisation approach was applied by \cite{whitelock2020disciplinary}, who grouped LMS activities into \textit{administration}, \textit{assessment}, \textit{content}, and \textit{engagement} categories, and by \cite{misiejuk2023changes}, who categorized LMS data to map \textit{administrative}, \textit{content}, and \textit{activity} variables. 

Beyond activity categorisation, other frameworks have proposed multilevel classifications distinguishing design decisions at the level of pedagogy, strategy, and individual task \cite{law2020multilevel}. Across the design analytics space more broadly, indicator development has focused predominantly on the student behavioural layer: the Constructivist Learning Design and Learning Analytics model \cite{banihashem2022impacts} maps LA indicators onto a seven-step constructivist design sequence, while other studies have mapped LMS activity logs onto LD taxonomy categories to examine student engagement patterns \cite{ahmad2024connecting}, or derived engagement indicators around constructs of workload and active learning \cite{ortiz2025extracting}. In each case, indicators characterise how students interact with a course rather than  properties of the design itself \cite{hernandez2019analytics}. Computable indicators that go beyond activity categorisation, capturing higher-order properties of LD directly from LMS data remain underdeveloped. This study addresses this gap by demonstrating a systematic process for deriving a set of composite LD indicators from Moodle LMS data, providing a scalable way to surface design choices and support the deliberate, evidence-informed practice required when instructors act as designers of technology-enhanced learning \cite{kirschner2015we}. 

\section{Methods}

Our dataset, exported from institutional Moodle, includes course LD data for all courses at a german large distance university that were active and visible to students in summer semester 2025 and winter semester 2025/2026. The dataset comprises 606 courses, with a mean of 7.3 sections ($SD = 5.7$) and a mean of 50.6 activities per course ($SD = 64.5$). The data include LD elements, but no user interaction logs. Basic preprocessing removed duplicate records, excluded courses that were not part of regular teaching, and dropped variables with substantial missing data or plugin-specific scope (Table \ref{tab:feature-definitions}). 

\begin{table}[htb]
\captionsetup{font=scriptsize}
\scriptsize
  \centering
  \caption{Overview of variables derived from Moodle data.}
  \label{tab:feature-definitions}
\begin{tabularx}{\linewidth}{>{\raggedright\arraybackslash}p{3.2cm}X}
    \toprule
    \textbf{Variable} & \textbf{Definition} \\
    \midrule
    \# of sections & Total number of sections in the course. \\
    \# of activities & Total number of activities in the course. \\
    \# of page items & Total number of page-type items in the course. \\
    \# of assessment activities & Total number of assessment-related activities. \\
    \# of receptive activities & Total number of activities focused on receptive learning \\ %(e.g., reading, watching) 
    \# of productive activities & Total number of activities focused on productive work 
    . \\ %(e.g., creating artifacts)
    \# of unique activity types & Total number of distinct activity types present in a course. \\
    assessment ratio & Proportion of graded activities out of all activities in a course.  \\
    nested surface ratio & Proportion of course-page activities that serve as containers linking to nested sub-pages or structures. \\
        collaboration ratio & Proportion of collaborative activities out of all activities in a course. \\ 
      LA block ratio & Proportion of LA-related blocks among all blocks present on course page.\\
mean orderliness & Mean regularity of activity-type sequencing within sections, averaged across sections (range: 0--1); computed as the ratio of distinct activity types to consecutive activity-type changes per section; higher values indicate more uniform, repetitive sequencing of activity types. \\
    section density & Mean \textit{\# of activities} per section in a course. \\
    SD activities & Standard deviation of \textit{\# of activities} per section in a course. \\
    \bottomrule
  \end{tabularx}
\end{table}

To determine the LD types in our dataset, addressing RQ1, we applied principal component analysis (PCA), a dimensionality reduction technique. PCA was applied to the standardised variable set, given that LMS-derived variables are interdependent, with instructors' integrated design choices affecting multiple structural variables at once, generating systematic co-variation that PCA reveals as interpretable latent dimensions. PCA was preferred over direct k-means on the full variable space as it reduces noise from correlated variables and produces a lower-dimensional representation that is more stable for clustering. While Singular Value Decomposition offers a related dimensionality reduction, PCA was chosen because it explicitly maximizes explained variance in the standardized variable space and produces components that are more interpretable in terms of the original LMS variables. As robustness checks, we repeated k-means and compared this solution with the PCA-based clustering using the Adjusted Rand Index \cite{hubert_comparing_1985}, and evaluated internal validity using average silhouette width, Calinski-Harabasz \cite{calinski_dendrite_1974}, and Davies-Bouldin indices \cite{davies_cluster_1979}. 

After determining the LD types, we characterised their distribution across three contextual dimensions: 
\textbf{Discipline}, categorised according to the university's faculty classification \textit{HASS} (Humanities and Social Sciences), \textit{STEM} (Mathematics and Computer Science), \textit{Psychology}, \textit{Commerce} (Business Administration and Economics), and \textit{Law}. \textbf{Activity load}, operationalised as a three-level band based on the number of activities per course: \textit{small} (fewer than 30), \textit{medium} (30--79), and \textit{large} (80 or more). 
This variable can serve as a proxy for the expected student workload and the granularity of course designs.
\textbf{Interface}, encompassing three Moodle-native course-page formats: \textit{one-topic}, a tabbed arrangement displaying one section at a time; \textit{tiles}, a card-grid arrangement presenting sections as visual tiles with optional progress indicators; and \textit{topics}, the default vertically stacked section arrangement.  Although interface choice is not inherently pedagogical, it is fundamental to the pedagogical interpretability of the course, as the visual organization of content can either make the intended learning path explicit or obscure it \cite{tabuenca_time_2015,shen_understanding_2020}.

For each dimension, we computed the proportional representation of each LD type and Fisher's exact test  (FET) after Bonferroni correction \cite{armstrong2014use,kim2017statistical}. In addition, we examined courses along three composite dimensions using scatter plots, and correlation within-type Pearson correlations between the two variables of each composite dimension to quantify the coherence of structural relationships per LD type. Additional ordinary least squares (OLS) regression lines were fitted within each LD type to visually characterise the direction and strength of these relationships in the scatter plots. PCA and k-means diagnostics are provided in the APPENDIX.

To address RQ2, focusing on how LD indicators can be derived from LMS course data and how they can be used to characterise the LD types, we followed a two-stage process. First, we derived three categorical measures from simple LMS variables. Thresholds for each measure were determined iteratively through expert discussion and evaluated against the observed empirical distributions in the RQ1 data, in order to ensure that the resulting bands reflected meaningful breaks rather than arbitrary cut-offs. Second, these categorical measures were combined with additional LMS variables to define three LD indicators, each targeting a different aspect of course design.

The thresholds for the indicators were again set through expert discussion and validated against the empirical distributions of specific variables in the Moodle data. To characterise the LD types, the indicators were then applied to the full Moodle dataset and their distributions compared across types. Finally, to evaluate the indicator's ability to differentiate meaningfully across LD types, we tested each indicator's distribution using FET with Monte Carlo approximation  ($B = 199,999$ resamples), which is appropriate for contingency tables containing small expected counts, and reported effect sizes as Cramér's V \cite{kim2017statistical}. Pairwise comparisons between LD types were Bonferroni-corrected within each indicator to control the family-wise error rate \cite{armstrong2014use}.

\section{Results}
\subsection{Identifying and Characterising LD Types in LMS Data} \label{rq1}

\begin{figure}[!htbp]
\centering
    \includegraphics[width=0.7\textwidth]{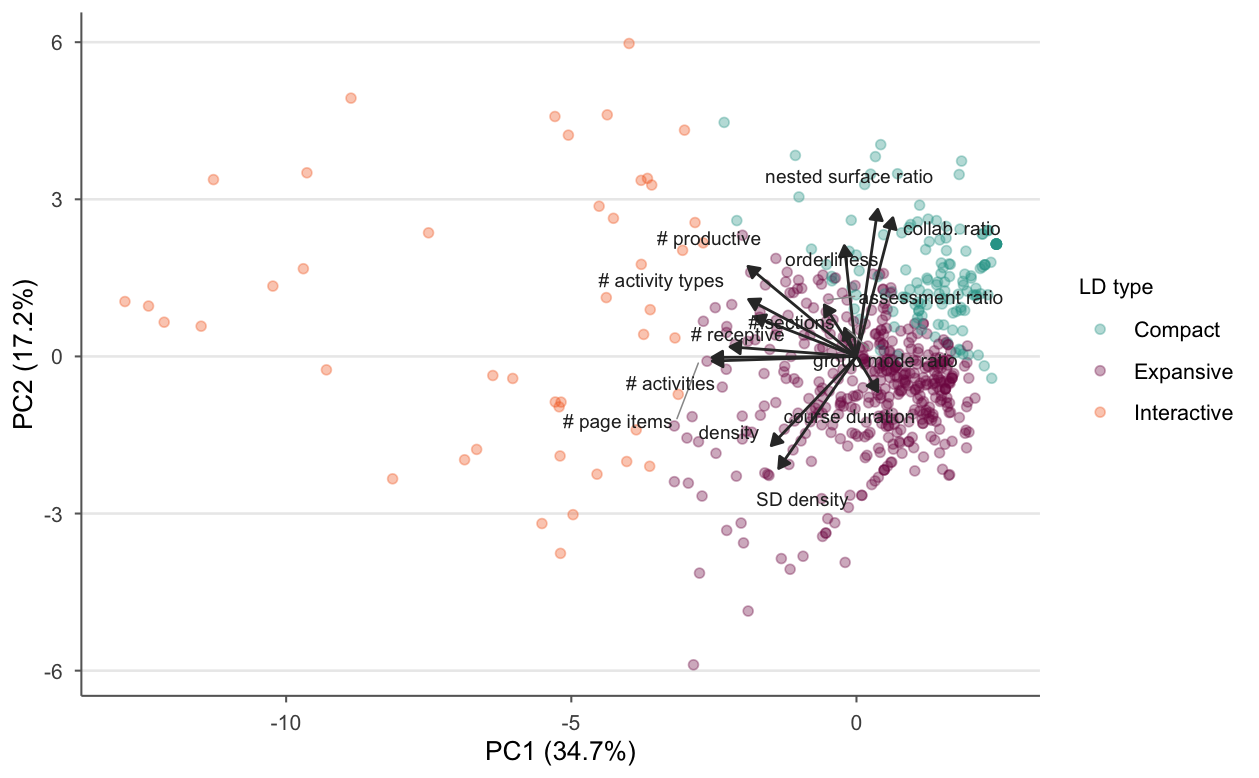}
    \caption{Courses (points, coloured by LD type) and variables (arrows) projected onto PC1 (34.7\,\%) and PC2 (17.2\,\%). Arrow direction indicates the orientation of each feature in the component space; arrow length reflects the strength of its contribution. Labels use shortened variable names for readability.}
    \label{fig:appendix-biplot}
\end{figure}

\begin{figure}[!htbp]
\centering
    \includegraphics[width=0.7\textwidth]{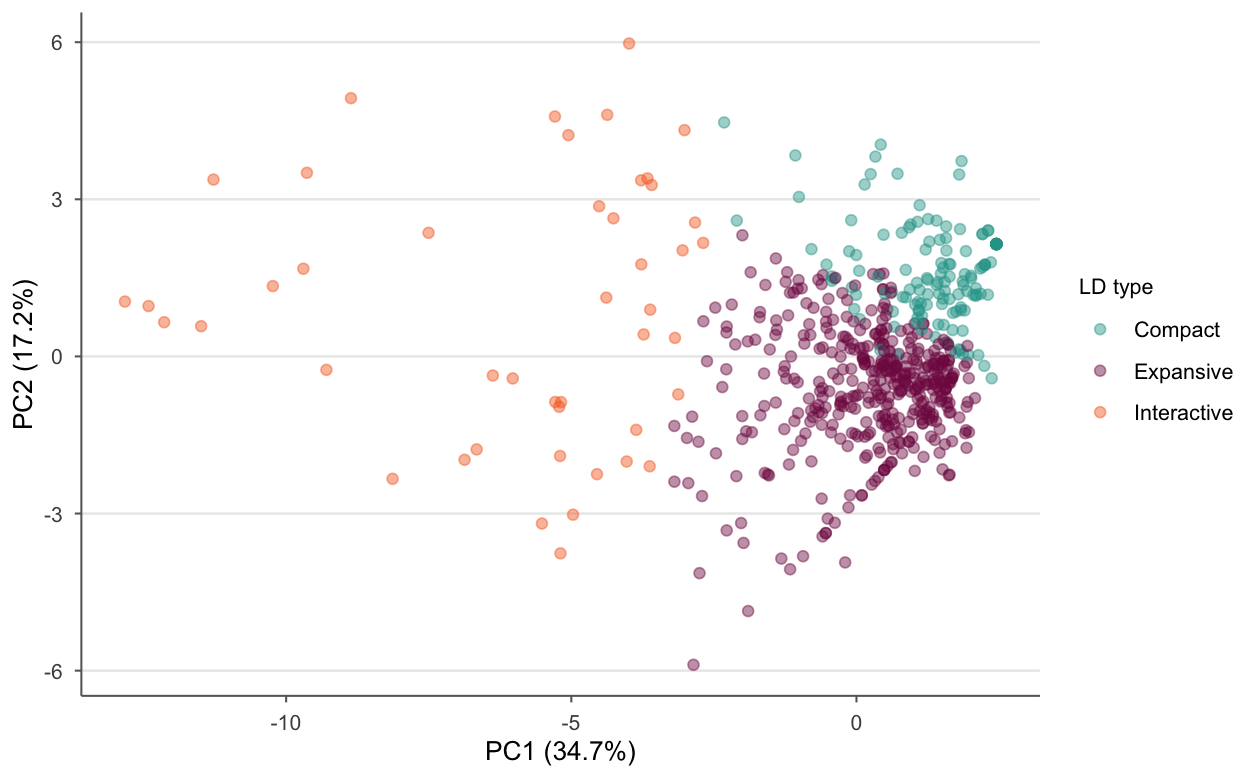}
    \caption{Course scores in PCA space by LD type. Each point represents one course projected onto the first two principal components and coloured by LD type: \textit{Compact} (teal), \textit{Expansive} (purple), \textit{Interactive} (orange). The three types form partially overlapping but distinguishable regions.}
    \label{fig:appendix-pca}
\end{figure}

To identify recurring LD types within the LMS data (RQ1), we applied PCA to the standardised variable set, retaining the leading components explaining approximately 92\,\% of the total variance. We then ran k-means clustering with Euclidean distance on the PCA scores. To select an appropriate number of LD types, we first examined the elbow and silhouette plots across candidate $k$ values. 
Although the silhouette score for k=3 (0.19) was only somewhat lower than for k=2 (0.30), we retained three clusters because they yielded more interpretable and pedagogically distinct LD types. The moderate silhouette scores suggest that courses form a continuum rather than discrete clusters; the three-cluster solution thus represents a pragmatic segmentation rather than naturally bounded types.
To inspect the structure of the three LD types, we projected courses onto the first two principal components and coloured points by LD type (APPENDIX Fig. \ref{fig:appendix-biplot}, \ref{fig:appendix-pca}). PC1 and PC2 together explained approximately 52\% of the variance (PC1: 34.7\%, PC2: 17.2\%), revealing three partially overlapping but clearly distinguishable clouds. PC1 functioned primarily as a course-scale axis, loading most strongly on total activities, number of sections, and page-based content. PC2 captured structural clarity and activity mix, combining positive loadings for orderliness, nested-surface ratio, and collaborative activities with negative loadings for section density and its variability. Thus, higher PC2 scores characterised more orderly and interaction-rich courses with more evenly distributed section layouts. 

One LD type extended furthest along PC1, corresponding to larger, more activity-dense courses, while the remaining two types were differentiated primarily along PC2. The names assigned to each type reflect their defining characteristics as revealed by the PCA structure:
\textbf{Expansive} LD type ($n=425$ courses) was characterised by high scores on PC1, reflecting courses with a large number of activities, many sections, and substantial page-based content. These are courses that cover broad ground through sheer volume of designed learning experiences. 
\textbf{Compact} LD type ($n=135$ courses) occupied a distinct region along PC2, distinguished by structural regularity and orderliness: courses with evenly distributed section layouts, a high nested-surface ratio, and low variability in section density. These are tightly organised courses that achieve coherence through deliberate structure rather than volume. 
\textbf{Interactive} LD type ($n=46$ courses) was set apart not by its activity load but by its activity composition. These courses show elevated collaborative activities and a higher proportion of interaction-oriented content, positioned at moderate PC1 scores but high on the PC2, the interaction-rich dimension.

\begin{figure}[htb]
    \centering
    \includegraphics[width=1\linewidth]{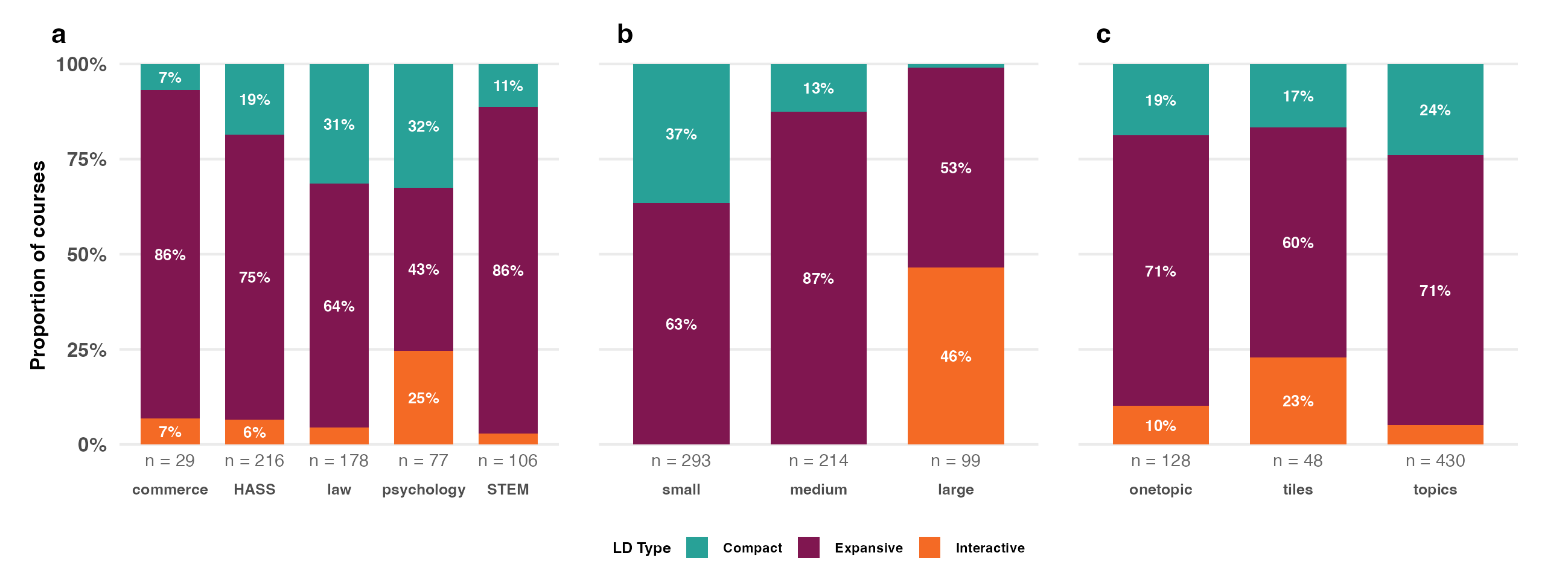}
    \captionsetup{font=scriptsize}
    \caption{Distribution of LD types by: (a) discipline, (b) activity load, (c) interface.}
    \label{fig:indicators:simple}
\end{figure}

Across all disciplines, \textit{Expansive} courses were the dominant pattern, \textit{Compact} courses formed a consistent secondary presence, and \textit{Interactive} courses were comparatively rare (Fig. \ref{fig:indicators:simple}a). However, the degree of concentration varied meaningfully across disciplines. HASS, Commerce, and STEM, shared a broadly similar profile dominated by \textit{Expansive} designs, ranging from approximately 75\% in HASS to 86\% in Commerce and STEM, with no significant differences among them after Bonferroni correction (Table \ref{tab:cluster-discipline}).
Psychology stood apart as the only discipline with a more balanced distribution and the only one in which \textit{	\textit{Interactive}} courses formed a substantial minority (24.7\%), differing significantly from all other disciplines (HASS: $p = .001$; Commerce: $p = .001$; STEM: $p = .001$; Law: $p = .001$). A further significant difference emerged between STEM and Law ($p = .002$), reflecting STEM's stronger concentration in \textit{Expansive} designs relative to Law's larger share of \textit{Compact} courses. Overall, FET confirmed that the distribution of LD types differed significantly across disciplines ($p < .001$, Cram\'er's $V = 0.24$).

\begin{table}[htb]
\captionsetup{font=scriptsize}
  \centering
  \caption{Distribution of LD types by discipline, activity load, and interface, with FET results.}
  \label{tab:cluster-discipline}
  \scriptsize
  \begin{tabular}{p{2.5cm}p{0.7cm}p{0.7cm}p{0.7cm}p{0.7cm}p{0.7cm}p{0.7cm}p{2.5cm}}
    \toprule
    & \multicolumn{2}{c}{\textit{Compact}} &
      \multicolumn{2}{c}{\textit{Expansive}} &
      \multicolumn{2}{c}{\textit{Interactive}} & \\
    \cmidrule(lr){2-3} \cmidrule(lr){4-5} \cmidrule(lr){6-7}
    Category & $n$ & \% & $n$ & \% & $n$ & \% & FET \\
    \midrule
    \multicolumn{7}{l}{\textit{Discipline}} & $p<.001$, $V=.24$\\
    \quad HASS        &  40 & 18.5 & 162 & 75.0 & 14 &  6.5 & \\
    \quad STEM        &  12 & 11.3 &  91 & 85.8 &  3 &  2.8 & \\
    \quad Commerce    &   2 &  6.9 &  25 & 86.2 &  2 &  6.9 & \\
    \quad Law         &  56 & 31.5 & 114 & 64.0 &  8 &  4.5 & \\
    \quad Psychology  &  25 & 32.5 &  33 & 42.9 & 19 & 24.7 & \\
    \midrule
    \multicolumn{7}{l}{\textit{Activity load}} & $p<.001$, $V=.50$ \\
    \quad Small       & 107 & 36.5 & 186 & 63.5 &  0 &  0.0 & \\
    \quad Medium      &  27 & 12.6 & 187 & 87.4 &  0 &  0.0 & \\
    \quad Large       &   1 &  1.0 &  52 & 52.5 & 46 & 46.5 & \\
    \midrule
    \multicolumn{7}{l}{\textit{Interface}} & $p<.001$, $V=.13$ \\
    \quad One-topic   &  24 & 18.8 &  91 & 71.1 & 13 & 10.2 & \\
    \quad Tiles       &   8 & 16.7 &  29 & 60.4 & 11 & 22.9 & \\
    \quad Topics      & 103 & 24.0 & 305 & 70.9 & 22 &  5.1 & \\
    \bottomrule
  \end{tabular}
\end{table}

Activity load was strongly related to LD type (Fig.\ref{fig:indicators:simple}b). Here, small, medium, and large refer to the scale of the Moodle course structure measured by number of activities, not student enrolment. \textit{Compact} LDs were almost exclusively small or medium in scale, with only a single course classified as large. \textit{Expansive} LDs spanned the full range of activity loads, from small to large, reflecting their position as the dominant pattern across the dataset (Table \ref{tab:cluster-discipline}). \textit{Interactive} LDs, by contrast, were found exclusively among large courses, marking activity volume as a defining variable of this type. FET confirmed that LD types differed significantly in their distribution across activity load bands ($p < .001$, Cram\'er's $V = 0.50$, a large effect), and Bonferroni-corrected pairwise comparisons showed that all three pairs differed significantly ($p_{\text{adj}} < .001$).

The distribution of interfaces across the three LD types was broadly similar: all three types relied predominantly on the topics arrangement (around 60--80\%), with smaller shares using one-topic and tiles (Table \ref{tab:cluster-discipline}). FET indicated a statistically significant difference ($p < .001$, Cram\'er's $V = 0.13$), but the effect was small and the differences were modest in practical terms. The clearest distinction was that \textit{Interactive} courses made comparatively greater use of the tiles arrangement than the other two LD types, and Bonferroni-corrected pairwise comparisons indicated that this difference was driven by the topics--tiles contrast ($p_{\text{adj}} < .001$), while one-topic did not differ significantly from either tiles ($p_{\text{adj}} = .334$) or topics ($p_{\text{adj}} = .253$). Overall, interface  was a weaker discriminator of LD type than discipline or activity load.

\begin{figure}
    \centering
    \includegraphics[width=1\linewidth]{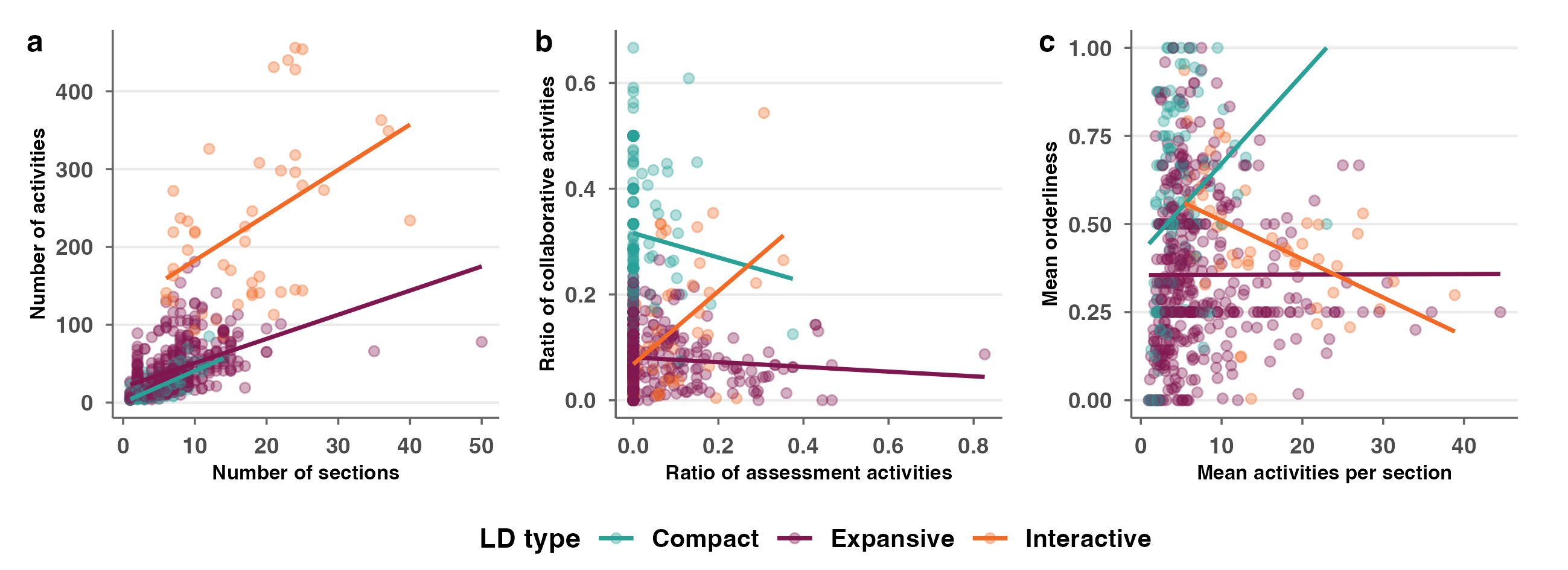}
    \captionsetup{font=scriptsize}
    \caption{Distribution of structural variables by LD type across three analytical dimensions: (a) \textit{\# of sections} $\times$ \textit{\# of activities}, (b) \textit{assessment ratio} $\times$ \textit{collaboration ratio}, and (c) \textit{section density} $\times$ \textit{mean orderliness}. Each point represents one course, coloured by LD type. Lines show within-type ordinary least squares (OLS) regression fits.}
    \label{fig:measures-regressions}
\end{figure}

To examine how the three LD types differ beyond their cluster assignments, we visualised their distributions across three dimensions. The scatter plot of \textit{number of sections} by the \textit{number of activities}  showed that \textit{Compact} courses clustered compactly in the low-scale region, combining few sections with few activities (Fig.\ref{fig:measures-regressions}a). \textit{Expansive} courses went beyond this dimension towards higher section and activity counts, while \textit{Interactive} courses occupied a distinct position, combining few sections with substantially more activities per section. Within-type correlations between sections and activities were strongest for \textit{Compact} ($r = .77$), and similarly moderate for both \textit{Expansive} and \textit{Interactive} ($r = .47$ each), though for structurally different reasons. The high correlation in \textit{Compact} courses reflects their tight, proportional scaling: few sections and few activities co-vary closely within a narrow range. The moderate correlation in \textit{Expansive}courses reflects greater heterogeneity across a wide scale range. For \textit{Interactive} courses, the same moderate correlation masks a structurally distinct pattern: sections and activities do not scale together but are instead decoupled, with activity load concentrated into comparatively few sections.

In the scatter plot showing the relationship between \textit{assessment ratio} and \textit{collaboration ratio}, \textit{Expansive} courses remained comparatively content-oriented, with lower shares of both activity types, reflecting their overall activity volume. \textit{Compact} courses showed the highest ratio of collaborative activities, with minimal use of assessment activities (Fig.\ref{fig:measures-regressions}b). \textit{Interactive} courses devoted a larger proportion of activities to collaborative and interactive elements, with moderate assessment use, suggesting a pedagogical orientation distinct from the other two types. Within-type correlations between assessment and collaboration ratios were moderate and positive for \textit{Interactive} courses ($r = .48$), indicating that these courses tend to combine higher collaboration with higher assessment use simultaneously, and near zero for both \textit{Compact} ($r = -.08$) and \textit{Expansive} courses ($r = -.08$).

The \textit{section density} by \textit{mean orderliness} scatter plot revealed that \textit{Compact} courses formed sparse but highly orderly layouts, combining low activity density per section with high structural consistency (Fig.\ref{fig:measures-regressions}c). \textit{Expansive} courses occupied a broad intermediate band, showing moderate density and variable orderliness. \textit{Interactive} courses tended toward denser sections with comparatively lower orderliness, suggesting that their activity concentration comes at the cost of structural regularity. Within-type correlations between mean activities per section and mean orderliness were weakly positive for \textit{Compact} ($r = .27$), near zero for \textit{Expansive} ($r = .00$), and moderately negative for \textit{Interactive} courses ($r = -.48$), confirming that within this type, higher section density is associated with lower structural regularity and concentration of activities into fewer sections comes at a measurable cost to layout consistency. 

\subsection{Deriving and Evaluating LD Indicators using LMS Data} \label{rq2}

To develop LD indicators from Moodle data (RQ2), we first operationalised three categorical measures that translated simple LMS variables into interpretable bands: \textit{assessment diversity}, capturing the volume and variety of assessment activities; \textit{section density}, classifying the mean number of activities per section; and \textit{orderliness}, reflecting the structural regularity of section layouts (Table \ref{tab:rq3-indicator-definitions}). 

\begin{table}[hbt]
\captionsetup{font=scriptsize}
\scriptsize
  \centering
  \caption{Derived categorical measures and LD indicators.}
  \label{tab:rq3-indicator-definitions}
  \begin{tabularx}{\linewidth}{p{2.5cm}X}
    \toprule
    Name & Thresholds \\
    \midrule
    \multicolumn{2}{l}{\textit{Categorical measures}} \\[-2pt]
    \multirow{4}{=}{\centering \textbf{Assessment diversity}} &
      none: \textit{\# of assessment activities} $= 0$ \\
      & 1--3 items: $1 \le$ \textit{\# of assessment activities} $\le 3$ \\
      & 4--10 items: $4 \le$ \textit{\# of assessment activities} $\le 10$ \\
      & 11+ items: \textit{\# of assessment activities} $\ge 11$ \\
    \midrule
    \multirow{3}{=}{\centering \textbf{Section density band}} &
      low: \textit{mean activities per section} $< 5$ \\
      & medium: $5 \le$ \textit{mean activities per section} $< 15$ \\
      & high: \textit{mean activities per section} $\ge 15$ \\
    \midrule
    \multirow{3}{=}{\centering \textbf{Orderliness band}} &
      low: \textit{mean orderliness} $< 0.33$ \\
      & medium: $0.33 \le$ \textit{mean orderliness} $< 0.66$ \\
      & high: \textit{mean orderliness} $\ge 0.66$ \\
    \midrule
    \multicolumn{2}{l}{\textit{LD indicators}} \\[-2pt]
    \multirow{4}{=}{\centering \textbf{Activity-mode balance}} &
      content-heavy: \textit{collaboration ratio} $< 0.05$ \& \textit{assessment ratio} $< 0.05$ \\
      & assessment-heavy: \textit{assessment ratio} $\ge 0.30$ \& \textit{collaboration ratio} $< 0.10$ \\
      & collaboration-heavy: \textit{collaboration ratio} $\ge 0.20$ \& \textit{assessment ratio} $\ge 0.20$ \\
      & mixed: all remaining combinations of collaboration ratio and assessment ratio \\
    \midrule
    \multirow{3}{=}{\centering \textbf{Structural clarity}} &
      clear: low/medium \textit{section density band} + high \textit{orderliness band} \\
      & cluttered: high \textit{section density band} + low/medium \textit{orderliness band} \\
      & mixed: all remaining combinations \\
    \midrule
\multirow{3}{=}{\centering \textbf{Analytics readiness}} &
  low: \textit{LA block ratio} $= 0$ \& \textit{number of nested items} $< 10$ \\
  & moderate:  \textit{LA block ratio} $> 0$ or 
    \textit{number of nested items} $\ge 10$ \\
  & high: \textit{LA block ratio} $> 0$ \& \textit{number of nested items} $\ge 100$ \\
    \bottomrule
  \end{tabularx}
\end{table}

The indicators were designed to capture three aspects repeatedly identified as important in the LA and design analytics literature: the pedagogical balance of activities, the structural organisation of the course, and the readiness of the design for analytics \cite{albo_knowledge-based_2022,ahmad2024connecting}. Because no established cut-offs exist for these aspects \cite{claassen_data_2025}, the thresholds themselves were determined iteratively through expert discussion and validation with empirical distributions. With the indicators and thresholds defined, we applied them to the full Moodle dataset to characterise the three LD types in greater detail. To evaluate whether the indicators differentiated meaningfully across LD types, we additionally tested their distributions for statistical differences.

\begin{figure} [hbt]
    \centering
    \includegraphics[width=1\linewidth]{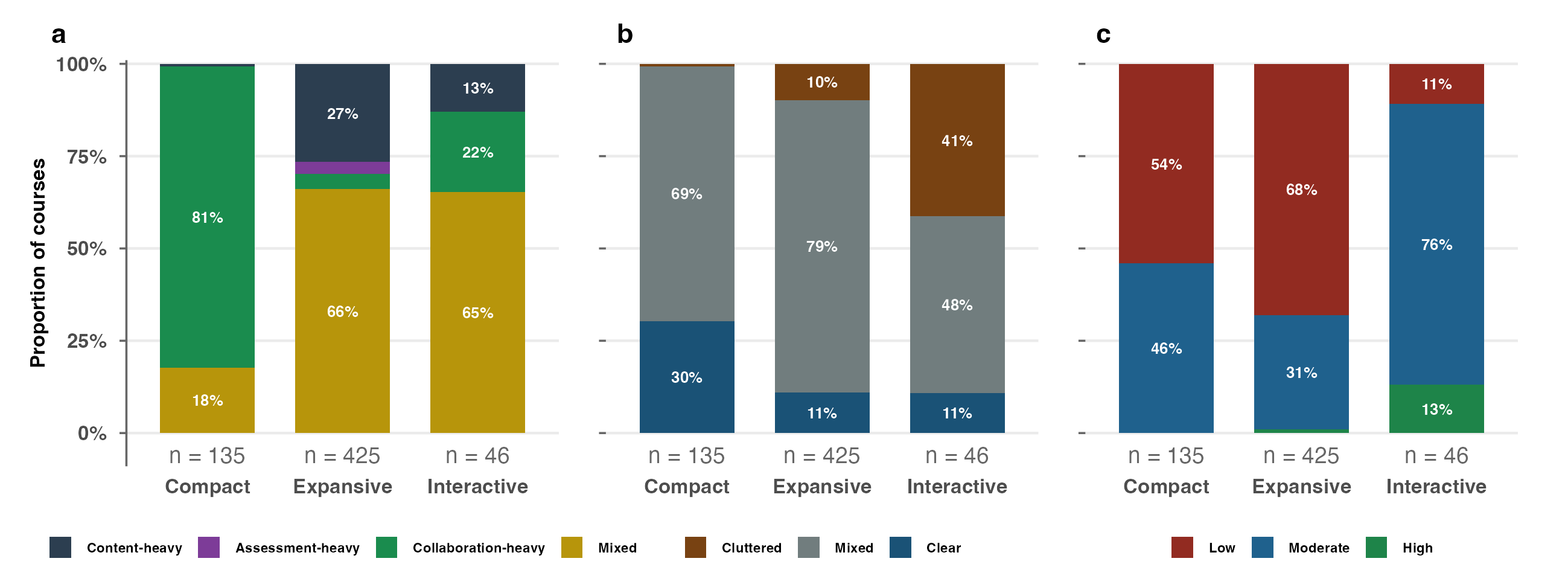}
    \captionsetup{font=scriptsize}
    \caption{Distribution of LD indicators by LD type for: (a) activity-mode balance, (b) structural clarity, (c) analytics readiness.}
    \label{fig:indicators}
\end{figure}

\textit{Activity-mode balance} is an indicator that classifies courses into four profiles according to the relative weight of content, assessment, and collaborative activities. It captures the degree to which a course concentrates learning opportunities within a single activity mode or distributes them across multiple modes. The statistical analysis revealed that \textit{activity-mode balance} differentiated significantly across the three LD types ($p < .001$, Cram\'er's $V = 0.26$), and Bonferroni-corrected pairwise comparisons confirmed that every pair of LD types was distinguishable (all $p_{\text{adj}} < .01$) (Table \ref{tab:rq2-combined}). 
The dominant pattern was the near-absence of content-heavy designs among \textit{Compact} LDs (0.7\%), in sharp contrast with \textit{Expansive} (26.6\%) and \textit{Interactive} (13.0\%) LDs (Fig. \ref{fig:indicators}a). 
\textit{Compact} LDs were characterised by a substantially higher share of collaboration-heavy designs (81.5\%) compared with \textit{Expansive} (4.0\%) and \textit{Interactive} (21.7\%) LDs. 
Conversely, mixed designs dominated \textit{Expansive} (66.1\%) and \textit{Interactive} LDs (65.2\%), compared with only 17.8\% in \textit{Compact} LDs. The strongest pairwise separation was between \textit{Compact} and \textit{Expansive} ($V = 0.35$).
\begin{table}[htb]
\captionsetup{font=scriptsize}
  \centering
  \caption{LD indicators by LD type, with FET results.}
  \label{tab:rq2-combined}
  \scriptsize
  \begin{tabular}{p{3.5cm}p{0.7cm}p{0.7cm}p{0.7cm}p{0.7cm}p{0.7cm}p{0.7cm}p{2.5cm}}
    \toprule
    & \multicolumn{2}{c}{\textit{Compact}} &
      \multicolumn{2}{c}{\textit{Expansive}} &
      \multicolumn{2}{c}{\textit{Interactive}} & \\
    \cmidrule(lr){2-3} \cmidrule(lr){4-5} \cmidrule(lr){6-7}
    Indicator & $n$ & \% & $n$ & \% & $n$ & \% & FET \\
    \midrule
    \multicolumn{7}{l}{\textit{Activity-mode balance}} &  $p<.001$, $V=0.26$ \\
    \quad Content-heavy     &   1 &  0.7 & 113 & 26.6 &  6 & 13.0 &  \\
    \quad Assessment-heavy  &   0 &  0.0 &  14 &  3.3 &  0 &  0.0 & \\
    \quad Mixed &  24& 17.8&  17 &  4.0 & 10 & 21.7 & \\
    \quad Collaboration-heavy     & 110& 81.5& 281 & 66.1 & 30 & 65.2 & \\
    \midrule
    \multicolumn{7}{l}{\textit{Structural clarity}} & $p<.001$, $V=0.27$ \\
    \quad Clear             &  41 & 30.4 &  47 & 11.1 &  5 & 10.9 &  \\
    \quad Mixed             &  93 & 68.9 & 336 & 79.1 & 22 & 47.8 & \\
    \quad Cluttered         &   1 &  0.7 &  42 &  9.9 & 19 & 41.3 & \\
    \midrule
    \multicolumn{7}{l}{\textit{Analytics readiness}} & $p<.001$, $V=0.27$  \\
    \quad Low             &   73&  54.1&   289&  68.0&   5&  10.9& \multirow{3}{*}{} \\
    \quad Moderate            &   62&  45.9&  132&  31.1&  35&  76.1& \\
    \quad High         &    0&   0.0&   4&   0.9&  6&  13.0& \\
    \bottomrule
  \end{tabular}
\end{table}

\textit{Structural clarity} combines section density and orderliness to characterise the course interface based on the relative weights. It reflects the extent to which the organisation of content into sections is consistent, predictable, and transparent to students. \textit{Structural clarity} also differentiated significantly across the three LD types ($p < .001$, $V = 0.27$), with all pairwise comparisons remaining significant after Bonferroni correction (all $p_{\text{adj}} < .001$). The largest effect was observed between \textit{Compact} and \textit{Interactive} LDs ($V = 0.57$): \textit{Compact} LDs were almost never cluttered (0.7\%) and were the most likely to be classified as \textit{clear} (30.4\%), whereas \textit{Interactive} LDs were cluttered in 41.3\% of cases and clear in only 10.9\% (Fig. \ref{fig:indicators}b). \textit{Expansive} LDs sat between the two extremes, with the largest share of \textit{mixed} designs (79.1\%) and intermediate levels of both clear (11.1\%) and cluttered (9.9\%) structures.

Finally, \textit{analytics readiness} reflects the degree to which a course is instrumented for analytics data collection. It operationalises the structural preconditions for generating meaningful interaction traces and supporting evidence-based monitoring of student engagement. For \textit{analytics readiness}, significant differences were found across all three LD types ($p < .001$, $V = 0.27$; all pairwise $p_{\text{adj}} < .05$ after Bonferroni correction). The strongest separations involved \textit{Interactive} LDs, which stood out with a \textit{moderate} readiness level in 76.1\% of cases and were the only type with a substantial share of \textit{high} readiness (13.0\%), compared with \textit{Expansive} ($V = 0.40$) and \textit{Compact} ($V = 0.46$) LDs (Fig. \ref{fig:indicators}c). \textit{Expansive} LDs were predominantly classified as \textit{low} readiness (68.0\%), while \textit{Compact} LDs were more evenly split between \textit{low} (54.1\%) and \textit{moderate} (45.9\%) and contained no designs at the \textit{high} level. The difference between \textit{Compact} and \textit{Expansive} was statistically significant but substantially smaller in magnitude ($V = 0.14$).

\section{Discussions and Conclusion}

This study examined how LD can be operationalised at scale from Moodle data, addressing two research questions: what LD types characterise Moodle course designs and how they differ in their characteristics and institutional distribution (RQ1), and how LD indicators can be derived from LMS course data to characterise these types (RQ2). 

\subsubsection{What the data reveal about Moodle course designs}
The analysis identified three distinct LD types, \textit{Compact}, \textit{Expansive}, and \textit{Interactive}, that differ systematically along two principal dimensions: activity load and the combination of structural clarity with interaction richness. 

\textit{Expansive} LDs dominate the dataset and are the predominant pattern across HASS, STEM, and Commerce, reflecting courses that cover broad ground through the sheer volume of designed learning experiences. An alternative explanation is that these courses share a similar default way of being configured in Moodle, a phenomenon also observed in large-scale analyses of university courses where a majority were found to follow a traditional, repository-like model \cite{revuelta-dominguez_structural_2026}. \textit{Expansive} courses  are characterised by a large number of activities, many sections, and substantial page-based content, spread across a wide range of activity loads with variable orderliness, a finding consistent with large-scale institutional analyses that have also identified a dominant pattern of high-volume, content-focused course designs \cite{rienties2017review}. The indicator analysis corroborated and extended this picture: \textit{Expansive} courses are the only type with a substantial share of content-heavy designs, and most of them show a mixed structural layout, neither clearly organised nor cluttered. They were also low in analytics readiness, a profile consistent with their high volume and broad coverage but one that points to a missed opportunity for data-informed course design at scale.
 
\textit{Compact} LDs form a consistent secondary presence characterised by tight, proportional scaling between sections and activities and by high structural regularity, and are particularly common in Law and Psychology. The indicator analysis corroborated this by showing that \textit{Compact} courses are almost never cluttered and are the most likely of any type to be classified as clear, while also revealing a subtler pattern: despite their limited scale, they contain a comparatively high share of collaboration-heavy designs. Concurrently, they were limited in analytics readiness, suggesting that structural clarity does not automatically translate into analytics readiness. These findings align with previous research on usability of mobile learning applications, which recommends that to improve usability, interfaces should be kept simple and compact, for instance by segmenting long pages into smaller, more manageable chunks to improve navigation \cite{tabuenca_time_2015}.

\textit{Interactive} LDs, though comparatively rare, form a distinctive group defined not by activity load but by activity composition, combining elevated collaboration and assessment use with denser, less orderly section layouts, and are concentrated almost exclusively in Psychology and in large courses. This aligns with disciplinary characteristics of Psychology as a discipline that has a strong tradition of active, discussion-based, and lab-oriented instruction \cite{whitelock2020disciplinary}. The indicator analysis corroborated and extended this: \textit{Interactive} courses were predominantly classified as mixed but with a substantial share of cluttered layouts, confirming that their activity concentration comes at a measurable cost to layout consistency, which highlights the critical need to interpret analytics within the context of the specific learning design to make sense of student behaviour \cite{shen_understanding_2020}.
As shown in previous research, interactive courses concentrate many activities into fewer sections, which makes them structurally denser but also more likely to appear cluttered and to require more careful structuring if they are meant to support students' course navigation and analytics support \cite{shen_understanding_2020}. Despite this, they were the best instrumented for analytics and the only type with a substantial share of high-readiness designs, showing that analytics readiness can coexist with, and perhaps even follow from, more complex and interaction-dense course structures.

Importantly, the three types are not simply different points on a common continuum but reflect structurally distinct organising logics: \textit{Compact} courses scale sections and activities in tight proportion, \textit{Expansive }courses spread activities across many sections with variable orderliness, and \textit{Interactive} courses decouple the two, concentrating activity load into comparatively few sections. This is reflected most sharply in the contrast between \textit{Compact} and \textit{Interactive} LDs, which showed the largest separation of the entire analysis on structural clarity, capturing the fundamentally different logic of small, tightly organised courses and large, mixed activity-mode ones. LD at scale, then, is not a single spectrum from "simple to complex", but a space of qualitatively different approaches that different disciplines adopt in different proportions.
 
\subsubsection{What LD indicators from LMS data reveal about course designs}
Building on these findings, we introduced a novel, transparent approach to deriving LD indicators from LMS data, combining theoretical grounding with iterative, data-driven threshold setting. To our knowledge, this is also the first work to propose indicators for LD: rather than simply classifying individual activities by type, our indicators aggregate and combine multiple structural variables into higher-order constructs that capture how these variables interact. Three indicators were derived, targeting the pedagogical balance of activities (\textit{activity-mode balance}), the structural organisation of the course (\textit{structural clarity}), and the readiness of the design for analytics (\textit{analytics readiness}). All three indicators differentiated significantly across the LD types, with every pair of types distinguishable on every indicator. Importantly, the indicators do not simply redescribe the clustering solution in a different form but add interpretable, pedagogically meaningful dimensions to it, most notably analytics readiness, which does not enter the clustering solution directly but proves to be a strong discriminator once introduced. This convergence between the raw LMS variables and the categorical indicators strengthens confidence in the robustness of the three LD types and offers a lightweight, scalable way to characterise them without manual coding.

\subsubsection{Pedagogical Impact and Future Directions}
That the introduced indicators can be derived from LMS data alone, without access to pedagogical intent, reinforces the argument that LMS data can serve as a proxy for underlying LD, providing a foundation for design analytics that focus on the properties of the course as designed \cite{hernandez2019analytics}. 
However, LMS data captures only one layer of the picture: it shows how courses are organised, but not why particular choices were made. 
Furthermore, using a snapshot of course designs excludes their evolution over time and the connection to student activity and LOs. This limitation stems from the available data: student interaction logs were not included and LOs are not systematically captured in the LMS. Consequently, the analysis enables the description of course designs, but prevents conclusions about how the identified LD types affect student engagement or performance. A key avenue for future research is to link these LD types with student interaction data and LOs, ideally through longitudinal analysis, to examine their correlation with learner success. 
Defining thresholds for categorical measures and LD indicators through expert discussion and validation with empirical distributions  represents an approach tailored to a specific institutional context, which may limit its generalizability. Future work should therefore test the robustness of the chosen thresholds through sensitivity analysis, validate the indicators across additional institutions, LMS platforms, and disciplines, and complement them with instructor rationales to move from characterising LD toward understanding the reasoning behind the specific design choices.

In conclusion, this study showed how LD can be operationalised at scale from Moodle data, contributing a large-scale empirical characterisation of 606 courses, the identification of three distinct LD types (\textit{Compact}, \textit{Expansive}, and \textit{Interactive}), and a novel approach to deriving LD indicators directly from LMS data. 
The emergence of generative AI is reshaping what LD has to do: instructors are integrating new tools, rethinking assessment, and restructuring courses to assure the quality of student learning. These pressures place instructors more squarely than ever in the role of designers \cite{kirschner2015we}, yet the choices they make at the level of the course as a whole remain largely invisible at scale, leaving both researchers and institutions without a clear view of how course design is actually evolving under these demands. The approach introduced here begins to close that gap, offering a lightweight, scalable way to surface those choices and, in doing so, supporting instructors in their designer role. More broadly, the indicators developed here are a first step toward a more deliberate, evidence-informed conversation about what LD should look like in a rapidly changing educational landscape.

\begin{credits}
\subsubsection{\ackname} This research was supported by the project LEAD:FUH funded by the Stiftung Innovation in der Hochschullehre (1001-3223), and Center of Advanced Technology for Assisted Learning and Predictive Analytics (CATALPA) at FernUniversit\"at in Hagen, Germany. We thank the Moodle team, particularly Daniel Poggenpohl, and the faculties’ associate deans for enabling this analysis.
\end{credits}

\newpage

% - Bibliography -
\bibliographystyle{splncs04}
\bibliography{mybibliography}

\newpage

\section{APPENDIX}
\subsection{k-means clustering}

\begin{figure}[!htbp]
\centering
    \includegraphics[width=0.6\textwidth]{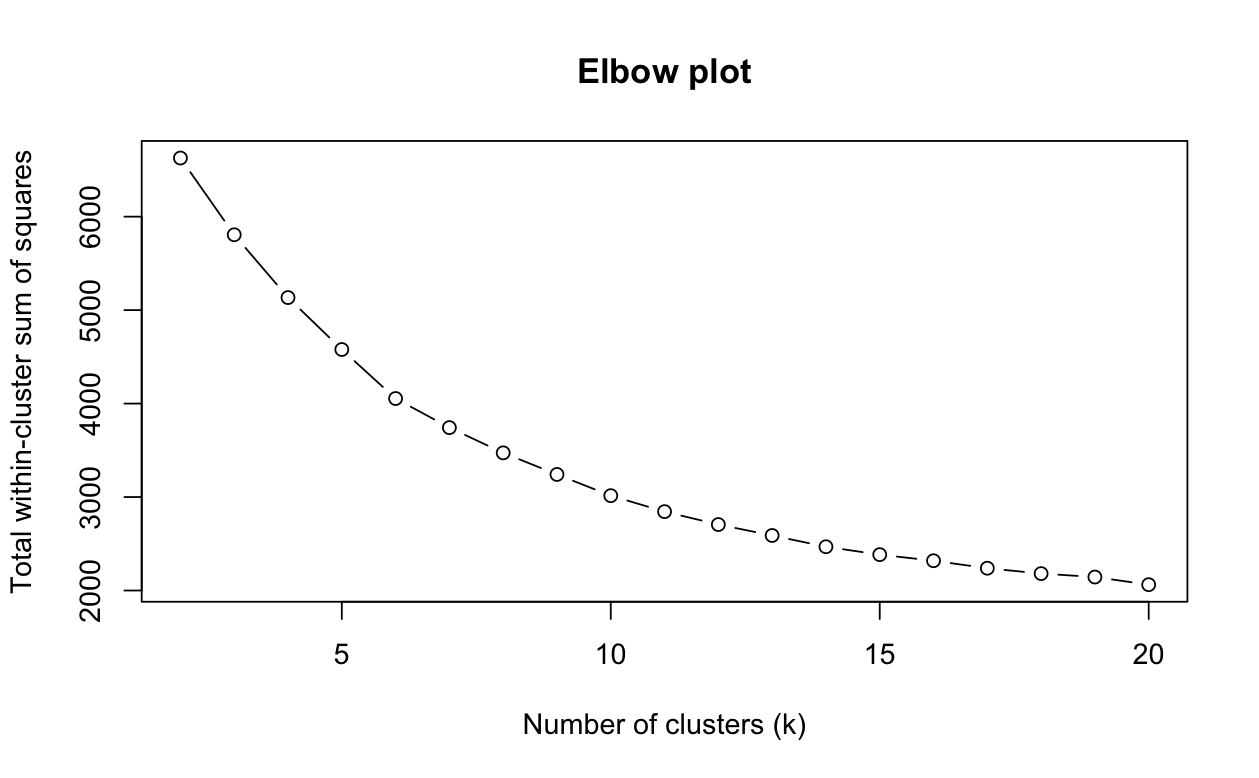}
    \caption{Elbow plot. Total within-cluster sum of squares plotted against 
    the number of clusters $k$ (1--20). The curve declines continuously without a sharp elbow, indicating no single dominant solution. $k = 3$ was selected based on convergent evidence from the silhouette plot and interpretability of the resulting cluster structure, balancing parsimony with meaningful differentiation of LD types.}
    \label{fig:appendix-elbow}
\end{figure}

\begin{figure}[!htbp]
\centering
    \includegraphics[width=0.7\textwidth]{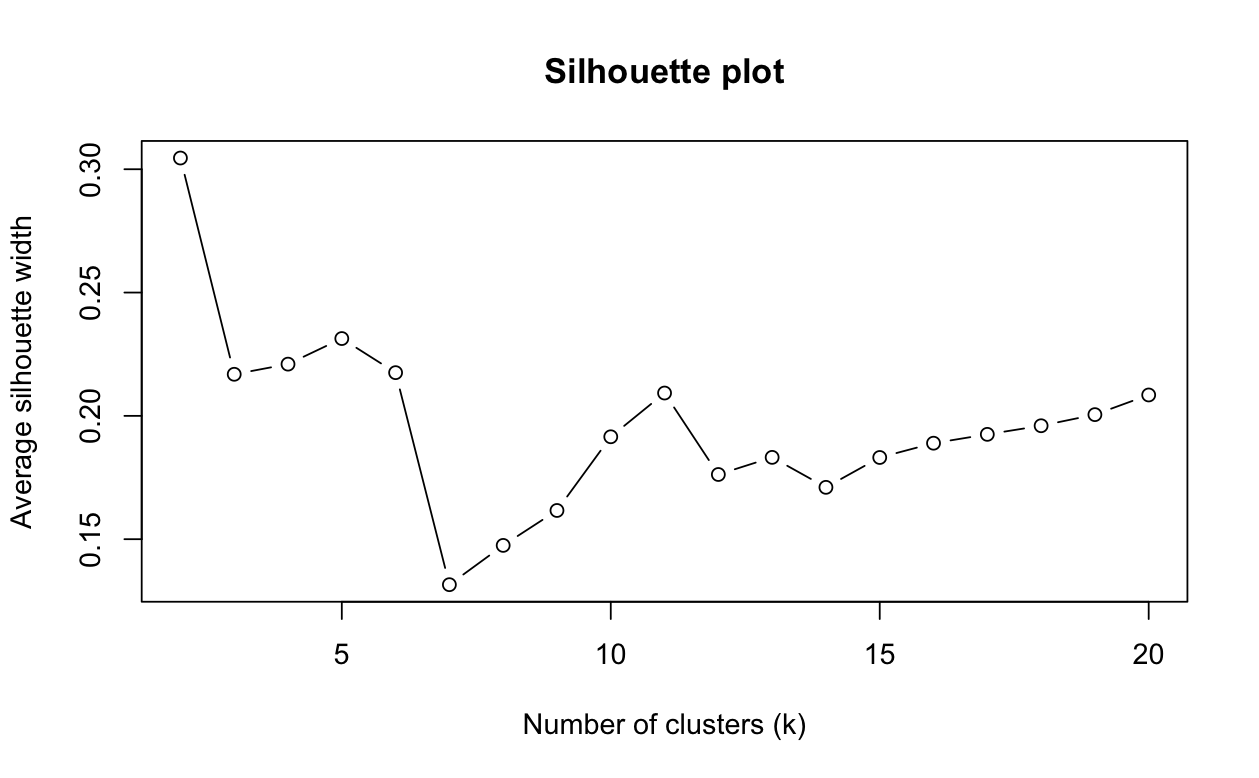}
    \caption{Silhouette plot. Average silhouette width across candidate values of $k$ (2--20). The highest silhouette is observed at $k = 2$ ($\approx 0.30$); $k = 3$ (silhouette $\approx 0.22$) was selected over $k = 2$ on grounds of interpretability, as it produces three structurally distinct and meaningfully differentiable LD types.}
    \label{fig:appendix-solhouette}
\end{figure}

\newpage

\subsection{Principal component analysis (PCA)}

\begin{figure}[!htbp]
\centering
    \includegraphics[width=0.7\textwidth]{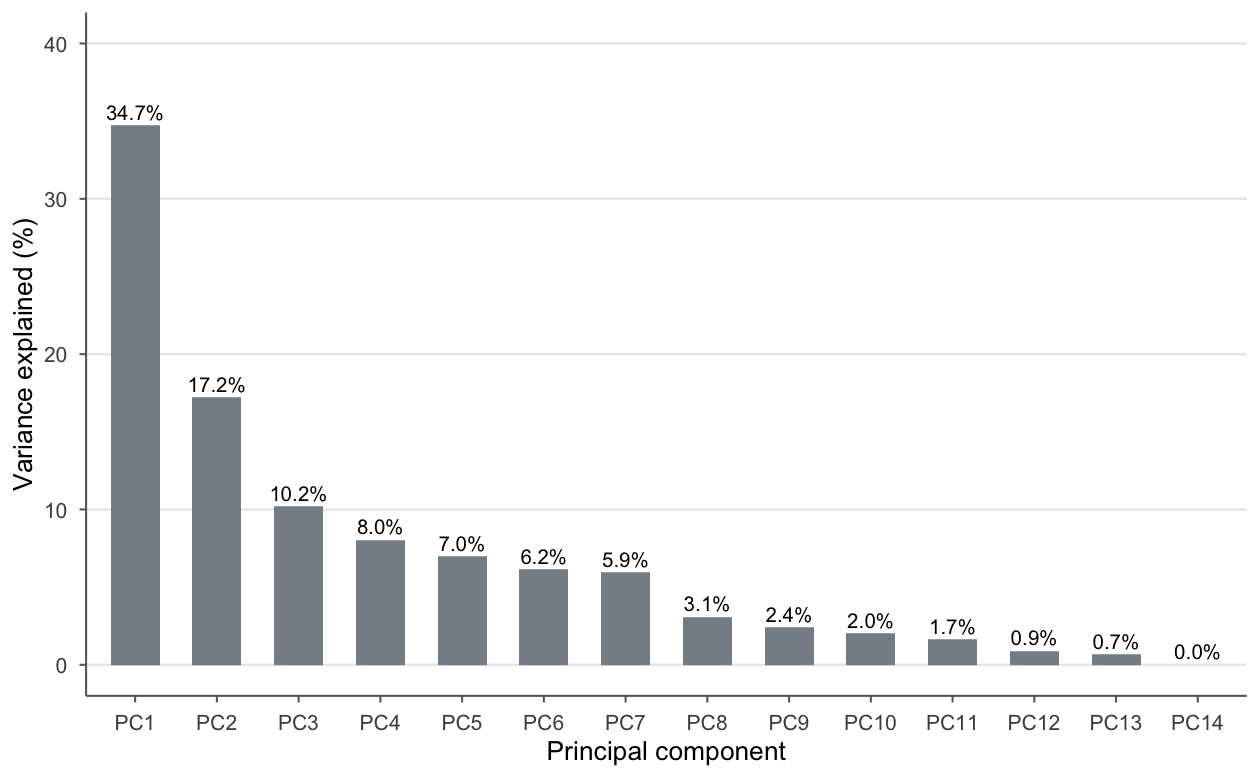}
    \caption{Scree plot. Percentage of variance explained by each of the 14 principal components. PC1 (34.7\,\%) and PC2 (17.2\,\%) together account for approximately 52\,\% of total variance; the leading components retained for clustering explain approximately 92\,\%.}
    \label{fig:appendix-scree}
\end{figure}

\begin{figure}[!htbp]
\centering
    \includegraphics[width=0.7\linewidth]{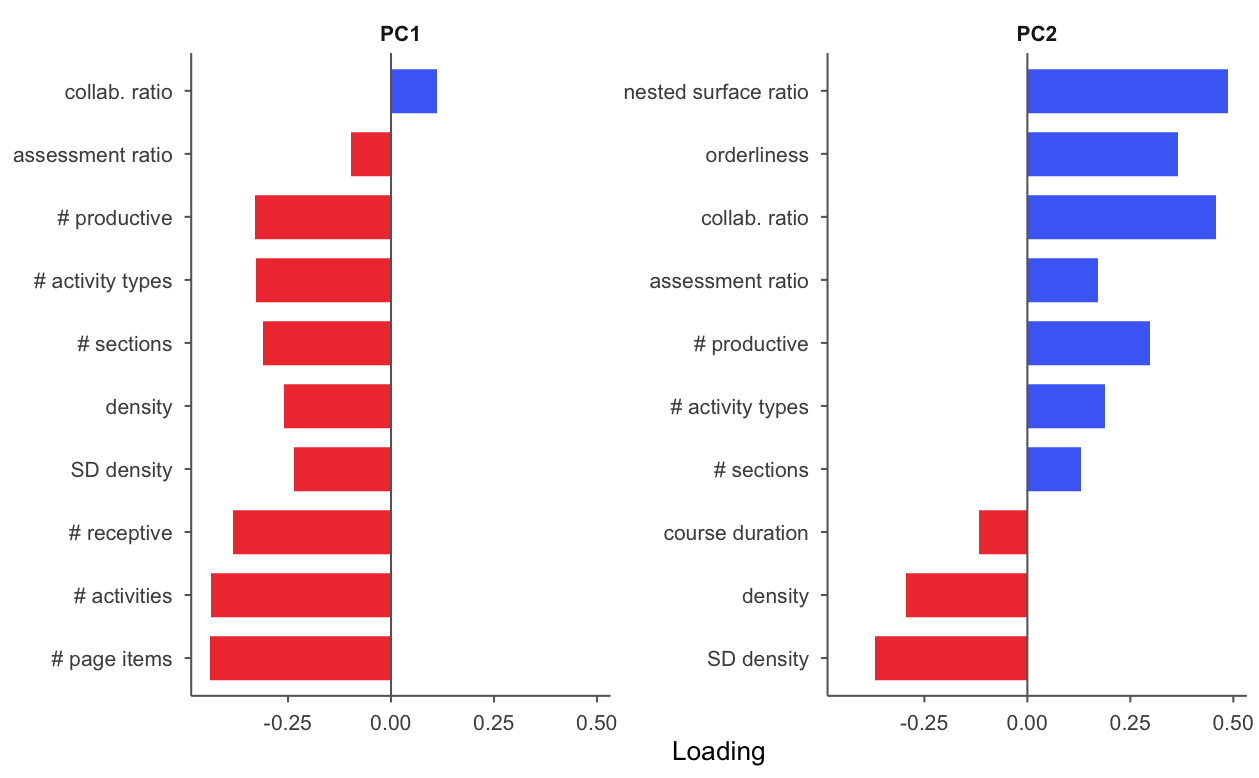}
    \caption{Top 10 absolute loadings for PC1 and PC2. The ten variables with the largest absolute loadings on each component are shown. Blue bars indicate positive loadings; red bars indicate negative loadings. PC1 is dominated by volume-related features (\# activities, \# page items); PC2 contrasts orderliness and nested-surface ratio against section density.}
    \label{fig:appendix-loadings}
\end{figure}

\end{document}